\documentclass[12pt]{article}
\usepackage{latexsym}
\usepackage{amssymb}
\usepackage{graphicx}
\def\de{\partial}

\def\K1{{\cal K}_{\bf 1}}
\def\Q20{{\cal Q}_{\bf 20'}}

\def\calK{{\cal K}}

\newbox\SlashedBox
\def\fs#1{\setbox\SlashedBox=\hbox{#1}
\hbox to 0pt{\hbox to 1\wd\SlashedBox{\hfil/\hfil}\hss}{#1}}
\def\hboxtosizeof#1#2{\setbox\SlashedBox=\hbox{#1}
\hbox to 1\wd\SlashedBox{#2}}

\def\ms#1{\setbox\SlashedBox=\hbox{$#1$}
\hbox to 0pt{\hbox to 1\wd\SlashedBox{\hfil/\hfil}\hss}#1}

\newcommand{\ie}{{\em i.e.~}}
\newcommand{\eg}{{\em e.g.~}}
\newcommand{\be}{\begin{equation}}
\newcommand{\ee}{\end{equation}}
\newcommand{\ba}{\begin{eqnarray}}
\newcommand{\ea}{\end{eqnarray}}
\begin{document}

\thispagestyle{empty}
\begin{flushright}
ROM2F/03/29
\end{flushright}
\vspace{1.2cm}
\begin{center}
{\Large {\bf Surprises from the Resolution \\
of Operator Mixing in ${\cal N} = 4$ SYM}}
\vspace{0.6cm}

{Massimo Bianchi,
Giancarlo Rossi and Yassen S. Stanev$^{\dagger}$} \\
\vspace{0.6cm}
{{\it Dipartimento di Fisica, \ Universit{\`a} di Roma \
``Tor Vergata''}} \\  {{\it I.N.F.N.\ -\ Sezione di Roma \
``Tor Vergata''}} \\ {{\it Via della Ricerca  Scientifica, 1}}
\\ {{\it 00133 \ Roma, \ ITALY}} \\
\end{center}
\vspace{0.6cm}

\setcounter{page}{1}

\begin{abstract}

We reexamine the problem of operator mixing in ${\cal N} = 4$ SYM.
Particular attention is paid to the correct definition of
composite gauge invariant local operators, which is necessary for
the computation of their anomalous dimensions beyond lowest
order. As an application we reconsider the case of operators with
naive dimension $\Delta_0=4$, already studied in the literature.
Stringent constraints from the resummation of logarithms in power
behaviours are exploited and the role of the generalized ${\cal N}
= 4$ Konishi anomaly in the mixing with operators involving
fermions is discussed. A general method for the explicit
(numerical) resolution of the operator mixing and the computation
of anomalous dimensions is proposed.

We then resolve the order $g^2$ mixing for the 15 (purely scalar)
singlet operators of naive dimension  $\Delta_0=6$. Rather
surprisingly we find one isolated operator which has a vanishing
anomalous dimension up to order $g^4$, belonging to an apparently
long multiplet. We also solve the order $g^2$ mixing for the 26
operators belonging to the representation ${\bf 20^\prime}$ of
$SU(4)$. We find an operator with the same one-loop anomalous
dimension as the Konishi multiplet.
\end{abstract}

\noindent
\rule{6.5cm}{0.4pt}

{\footnotesize ${}^{\dagger}$~On leave of absence from Institute for
Nuclear Research and Nuclear Energy, Bulgarian Academy of Sciences,
BG-1784, Sofia, Bulgaria}

\newpage
\section{Introduction and summary}
\label{intro}

In~\cite{Bianchi:2002rw} we resolved the mixing among scalar
primary operators of naive scale dimension $\Delta_{0} = 4$ in the
${\bf 20}^\prime$ representation of the $SU(4)$ R-symmetry of the
${\cal N}=4$ SYM theory and computed their anomalous dimensions at
order $g^{2}$. A similar analysis was independently performed by
the authors of ref.~\cite{Arutyunov:2002rs} in the $SU(4)$ singlet
sector. More involved mixing problems have been
studied~\cite{Kristjansen:2002bb,Constable:2002hw,Beisert:2002bb,Constable:2002vq}
in the BMN limit~\cite{Berenstein:2002jq} (large R-charge sector)
of ${\cal N}=4$ SYM, conjectured to be dual to type IIB
superstring on a
pp-wave~\cite{Blau:2001ne,Blau:2002dy,Blau:2002mw,Figueroa-O'Farrill:2002ft,
Blau:2003rt,Metsaev:2001bj,Metsaev:2002re,Metsaev:xc}. Although
the very issue of holography remains somewhat mysterious in this
setting, the authors of ref.~\cite{Santambrogio:2002sb} were able
to show that tree-level superstring predictions for the anomalous
dimensions of BMN operators with two ``impurities'' are consistent
at the planar level ($N_c>>1$, $g^2N_c={\rm{fixed}}$, $J^2\approx
N_c$, so that $g^2N_c/J^2<<1$) with field theoretic results.
Non-planar contributions that probe string interactions in the
pp-wave background are subtler and require analyzing the mixing of
operators with different number of
traces~\cite{Kristjansen:2002bb,Constable:2002hw,Beisert:2002bb,Constable:2002vq}
along the lines of~\cite{Bianchi:2002rw,Arutyunov:2002rs}. Some
degeneracies in the anomalous dimensions of ``renormalized'' BMN
operators with two impurities, that at first looked puzzling, have
been completely clarified in~\cite{Beisert:2002tn} on
supersymmetry grounds and the BMN mixing problem has been
rephrased in terms of a dilatation
operator~\cite{Beisert:2002ff,Kim:2003rz} and extensively
investigated~\cite{Chu:2002pd, Chu:2002qj, Chu:2002wj, Chu:2003ji,
Chu:nb, Gursoy:2002yy, Gursoy:2002fj,
Freedman:2003bh,Eden:2003sj,Klose:2003tw}.

A remarkable twist of the situation was brought about by Minahan
and Zarembo~\cite{Minahan:2002ve} who have shown, quite
independently of the BMN limit, that the one-loop dilatation
operator in the sector of purely scalar operators can be viewed as
the Hamiltonian of an integrable $SO(6)$ spin chain. Their results
have been extended in ref.~\cite{Beisert:2003tq}, where it was
shown that ``pure operators'' (\ie\ lowest scalar components of
naively 1/4 BPS multiplets, with $\Delta = 2k + l$, belonging to
the representation $[k,l,k]$ of the $SU(4)$ R-symmetry which can
only mix among themselves), are governed by an $SU(2)$ spin chain
which is integrable up to at least $(g^2)^3$ (three-loops).
Evidence that the planar dilatation operator for two-impurity BMN
operators is of the form predicted by string theory was
also given. Later on it was proven that the full one-loop
dilatation operator is the Hamiltonian of a super spin
chain~\cite{Beisert:2003jj,Beisert:2003yb} and some interesting
``closed sectors'' have been identified and analyzed in connection
with higher spin symmetry enhancement~\cite{Bianchi:2003wx,
Beisert:2003te}, producing an overwhelming set of data that expose
some intriguing regularity~\footnote{In particular the ``golden
ratio'' found in~\cite{Bianchi:2002rw} reappears several times
in~\cite{Beisert:2003jj,Beisert:2003yb}.}. Several issues related
to the question of integrability beyond one-loop have been
recently addressed. In particular in~\cite{Beisert:2003ys} it has
been clarified how to accommodate mixing of operators with
different ``length'' (\ie\ number of constituents) in the spin
chain picture and incorporate the presence of terms with odd
powers of $g$ in the dilatation operator.

Our aim in this note is to shed some light on certain
field-theoretical issues that seem to have become topical and put
forward few interesting new results that deserve a deeper
understanding. In particular, in Section~\ref{composite} we
reexamine the general problem of operator mixing in ${\cal N} = 4$
SYM, paying special attention to the correct definition of
composite gauge invariant local operators. This theoretical step
is crucial for the computation of anomalous dimensions and to make
sound statements about integrability at order higher than
$g^2$~\cite{Beisert:2003ys, Klose:2003qc}. As an application, we
reconsider the case of operators belonging to the irrep ${\bf
20}^\prime$ of $SU(4)$ with naive dimension $\Delta_0=4$, already
studied in the literature~\cite{Bianchi:2002rw}. In
Section~\ref{skonan}, we discuss the role of the generalized
${\cal N} = 4$ Konishi anomaly~\cite{Konishi:1983hf, Amati:ft,
Cachazo:2002ry} in the mixing pattern of operators with fermion
impurities~\cite{Eden:2003sj}. In Section~\ref{opemixres}
stringent constraints from the resummation of logarithms in
(conformal) power-like terms are exploited to give a general
method for the explicit (perturbative) resolution of operator
mixing and the calculation of anomalous dimensions. In
Section~\ref{anodim6} we resolve the mixing of the 15 (purely
scalar) singlet operators of naive dimension $\Delta_0=6$ at order
$g^2$. Rather surprisingly we find one operator ${\cal T}$ that,
despite the fact that it is the lowest component of a long
multiplet, has vanishing anomalous dimension up to order $g^4$.
 A similar analysis carried out up to order $g^2$ aimed at
 resolving
the mixing of the 26 operators with naive dimension $\Delta_0=6$
in the ${\bf 20^\prime}$ of $SU(4)$ shows the existence of one
operator with the same one-loop anomalous dimension as the Konishi
multiplet.   In Section~\ref{concsumm} we draw our conclusions.

\section{The definition of the composite operators}
\label{composite}

In this section we present a general discussion of the problem of
constructing renormalizable gauge invariant composite operators in
a (super) conformal theory, \ie\ operators that, besides being
multiplicative renormalizable (m.r.) have well defined
transformation properties under dilation, actually under the whole
(super) conformal group.

We shall illustrate this issue in the simple context of operators
of naive dimension $\Delta_0=4$ in the ${\bf 20}^\prime$ irrep of
$SU(4)$. There are 4 purely scalar operators of this kind, which
belong, at least at tree-level, to semishort
multiplets~\cite{Andrianopoli:1999vr, Andrianopoli:1998ut,
Andrianopoli:1998nc,Andrianopoli:1998jh, Dolan:2002zh}. Three of
them have non-vanishing order $g^2$ anomalous dimension, while the
fourth operator, ${\cal D}^{ij}_{{\bf 20}^\prime}$ (whose
components we will denote for short ${\cal D}_{{\bf 20}^\prime}$
or ${\cal D}^{ij}$ depending on the context), has vanishing
anomalous dimension~\cite{Bianchi:2002rw, Arutyunov:2000ku,
Arutyunov:2001mh, Arutyunov:2000im, Eden:2000bk, Eden:2001ec,
Dolan:2001tt, Dolan:2000ut, Penati:2001sv}. As an illustration of
the problems one encounters we choose to discuss the simple case
of  ${\cal D}_{{\bf 20}^\prime}$. One might be tempted to write
${\cal D}_{{\bf 20}^\prime}$  in the form
\begin{equation}
{\cal D}^{ij}(x)=\ : \sum_{k=1,\ldots,6}\ {\cal Q}^{ik}(x){\cal Q}^{jk}(x)
-{\delta^{ij} \over 6}
\sum_{k,l=1,\ldots,6}{\cal Q}^{kl}(x) {\cal Q}^{kl}(x) \ : \, ,
\label{defD_1}
\end{equation}
where ${\cal Q}^{ij}_{{\bf 20}^\prime}$ are the lowest component scalars of
the ultrashort 1/2 BPS ${\cal N}=4$ supercurrent multiplet
\begin{equation}
{\cal Q}^{ij}(x)\ = \ : \ {\rm Tr}(\varphi^i(x)\varphi^j(x))-
{\delta^{ij}\over 6}
\sum_{k=1,\ldots,6} {\rm Tr}(\varphi^k(x) \varphi^k(x)) \ :  \, ,
\label{defQ_1}
\end{equation}
with $\varphi^i$, $i=1,\dots ,6$ the ${\bf 6}$ fundamental scalars
of ${\cal N}=4$ SYM. In both the above formulae the normal product
$ :: $ denotes as usual the omission of self-contractions
(contractions of fields sitting at the same point), which are
anyway absent in eq.~(\ref{defQ_1}). In Section~\ref{solut} we
will see how to give a precise definition of these operators in
the regularized theory.

\subsection{The problem}
\label{problem}

Suppose that we want to resolve the mixing problem of the set of
operators, ${\cal O}_p$, of the type described above (\ie\ with
naive dimension $\Delta_0=4$ and belonging to the ${\bf
20}^\prime$ irrep of $SU(4)$) among which we have ${\cal D}_{{\bf
20}^\prime}$. The final renormalized operators (which we shall
denote by $\widehat {\cal O}_p$) will have to have well defined
(anomalous) dimensions and hence will define an orthogonal basis
with respect to the inner product represented by the 2-point
functions $\langle{\cal\widehat O}_p^\dagger(x){\cal\widehat
O}_q(0)\rangle$. Furthermore ${\cal \widehat O}_p$ should have a
vanishing 2-point function with the protected chiral primary
operators (CPO's) ${\cal Q}^{ij}$ of eq.~(\ref{defQ_1}), which
have dimension $\Delta=2$. Thus we must have for any $p$
\begin{equation}
\langle \widehat {\cal O}^{ij}_p (x) \  {\cal Q}^{kl} (y)  \rangle \ = \ 0
\label{oq}
\end{equation}
to all orders in perturbation theory and non-perturbatively. On the other
hand general principles of Quantum Field Theory tell us that the final
renormalized operators, $\widehat{\cal O}_q$, must be linear
combinations of the bare ${\cal  O}_p$ operators. In formulae one
must have
\begin{equation}
\widehat {\cal O}_p = \sum_q Z_{pq} \ {\cal O}_q  \, ,
\label{ohato}
\end{equation}
where the mixing matrix, $Z$, is assumed to be an invertible
(not necessarily symmetric) matrix, so that one can invert the
relation~(\ref{ohato}), getting
\begin{equation}
{\cal O}_q = \sum_p Z^{-1}_{qp} \ \widehat {\cal O}_p  \, .
\label{oohat}
\end{equation}
A simple perturbative computation shows however that
${\cal D}_{{\bf 20}^\prime}$ of eq.~(\ref{defD_1}) fails to fulfill
condition~(\ref{oq}) already at order $g^2$, as an explicit computation gives
\begin{equation}
\langle  {\cal D}^{ij} (x)\ {\cal Q}^{kl} (y) \rangle \vert_{g^2} \neq 0 \, .
\label{DQ}
\end{equation}
As a result this operator cannot be expressed as suggested by
eq.~(\ref{oohat}). Similar considerations apply to essentially all
the naive definitions of ${\cal D}_{{\bf 20}^\prime}$ used in the literature,
despite the fact that it is the lowest component of an exactly semishort
multiplet~\cite{Andrianopoli:1999vr,Andrianopoli:1998ut,Andrianopoli:1998nc,Andrianopoli:1998jh,Dolan:2002zh}.

A careful inspection of the above argument shows that the problem
arises from the incompatibility of the (naive) definition of
${\cal D}_{{\bf 20}^\prime}$ given in eq.~(\ref{defD_1}) with the
assumption that the mixing matrix $Z_{pq}$~(\ref{ohato}) is
invertible and the lack of an explicit regulator in
eq.~(\ref{defD_1}). Notice that, on the contrary, protected
operators in short BPS multiplets, such as ${\cal Q}_{{\bf
20}^\prime}$ of eq.~(\ref{defQ_1}), do not seem to suffer of the
same problem.

In the next section we will show that, as soon as the theory has
been properly regularized, the problem of constructing m.r.
operators with well defined conformal dimension can be fully
solved, so that the vanishing of correlators such as $\langle {\cal
D}_{{\bf 20}^\prime} \ {\cal Q}_{{\bf 20}^\prime} \rangle$ will be
guaranteed.

\subsection{The solution}
\label{solut}

Let us regularize the theory, say, by point splitting. Given a bare
operator ${\cal O}_p$ we shall call $\widetilde {\cal O}_p$ the
subtracted operator which does not mix with any operator of
dimension smaller than the naive dimension of ${\cal O}_p$. In the
case of ${\cal D}_{{\bf 20}^\prime}$ we find for the regularized
operator, $\widetilde {\cal D}_{{\bf 20}^\prime}$,  the expression
\begin{eqnarray}
&&\widetilde {\cal D}^{ij}(x)  = \lim_{\epsilon \rightarrow 0} \left\{
{\cal Q}^{ik}(x+{\epsilon \over 2 }) {\cal Q}^{jk}(x-{\epsilon \over 2 })
-{\delta^{ij} \over 6 }
{\cal Q}^{kl}(x+{\epsilon \over 2 }) {\cal Q}^{kl}(x-{\epsilon \over 2 })\ +
\right.  \nonumber \\
&& \left.
-{5\over 3 \pi^2\epsilon^2}\left[{\rm Tr}\left(\varphi^i(x+{\epsilon\over 2})
\varphi^j(x-{\epsilon \over 2 })\right) -{\delta^{ij} \over 6}
{\rm Tr}\left(\varphi^k(x+{\epsilon \over 2})
\varphi^k(x-{\epsilon\over 2})\right)\right]\right\} \, .
\label{defD_2}
\end{eqnarray}
The second line in this equation can be expanded in a power series
of the parameter $\epsilon$ and can be simplified with the help of
the identity (valid after averaging over the angular dependence of
$\epsilon$)
\begin{equation}
\lim_{\epsilon \rightarrow 0}
{\epsilon_{\mu}\epsilon_{\nu}\over\epsilon^2}={\delta_{\mu \nu}\over 4}\, .
\label{EPS}
\end{equation}
After redistributing derivatives, we finally obtain
\begin{eqnarray}
\hspace{-1cm}&&\widetilde {\cal D}^{ij}(x) \ = \nonumber \\
\hspace{-1cm}&&
\lim_{\epsilon \rightarrow 0} \left\{
{\cal Q}^{ik}(x+{\epsilon \over 2 }) {\cal Q}^{jk}(x-{\epsilon \over 2 })
-{\delta^{ij} \over 6}
{\cal Q}^{kl}(x+{\epsilon \over 2 }) {\cal Q}^{kl}(x-{\epsilon \over 2 }) -
{5\over 3\pi^2\epsilon^2}{\cal Q}^{ij}(x) \right\}+
{5\over 96\pi^2}\Box{\cal Q}^{ij}(x)\nonumber \\
\hspace{-1cm}&& -  \ {5 \over 48 \pi^2}  \left[ {\rm Tr}
\left((\Box \varphi^i(x)) \varphi^j(x)\right)
+{\rm Tr}\left(\varphi^i(x) (\Box\varphi^j(x))\right)
-{\delta^{ij} \over 3 }
{\rm Tr}\left((\Box \varphi^k(x)) \varphi^k(x)\right) \right] \ .
\label{defD_3}
\end{eqnarray}
The two lines of this equation have rather different physical
content, so we shall comment on them separately. The first line is
just an explicit though cumbersome way to eliminate
self-contractions in the composite operator, \ie it implements the
normal ordering ($ :: $) as defined after eq.~(\ref{defQ_1}).
These terms are always present, even in the free field theory at
$g=0$. Their explicit form depends on the way the regulator is
introduced. For example, if we use an asymmetric point-splitting,
the $\Box{\cal Q}^{ij}$ term is replaced by
$(\epsilon\partial){\cal Q}^{ij}/\epsilon^2$. If instead of
point-splitting, we use dimensional regularization, the
quadratically divergent subtraction can be omitted. The situation
is different for the terms in the second line. They all contain
the $\Box\varphi^i(x)$ operator, which by the use of the field
equations gives rise to terms that are non-vanishing in the
interacting theory, and do not depend on the particular
regularization scheme adopted in the calculations. These terms,
which have no counterpart in the naive definition of
eq.~(\ref{defD_1}), are indeed necessary (and sufficient) to get
the vanishing of the three-point function of $\widetilde{\cal
D}^{ij}$ inserted with two fundamental fields $\varphi^k$, \ie
\begin{equation}
\langle \widetilde  {\cal D}^{ij} (x) \ \varphi^k(y_1) \
\varphi^l(y_2)
 \rangle \vert_{g^2} \ =  \ 0  \, ,
\label{Dphiphi}
\end{equation}
which in turn implies the correct vanishing of the two-point function
\begin{equation}
\langle \widetilde{\cal D}^{ij}(x)\ {\cal
Q}^{kl}(y)\rangle\vert_{g^2}\ = \ 0 \, .
\end{equation}
The result of this analysis is that the m.r.\ operator
$\widetilde{\cal D}^{ij}(x)$ has a hidden $g$ dependence from the terms
in the second line of the r.h.s.\ of eq.~(\ref{defD_3}). This $g$
dependence can be made explicit with the help of the field
equations, which read
\begin{equation}
D^2\varphi^i = \sqrt{2} g (\tau^i_{AB} [\lambda^A, \lambda^B] +
h.c.) + g^2 \ [\varphi^j,[\varphi^i, \varphi_j]] \, .
\label{eqmo}\end{equation} where $\lambda^A$, $A=1,\dots,4$ denote
the ${\bf 4}$ gaugini and $\tau^i_{AB}$ the $4\times 4$
(antisymmetric) chiral blocks of the $D=6$ $\gamma$-matrices.
Consequently we can write an expansion of the type
\begin{equation}
\widetilde {\cal D}^{ij}(x)\ = \ {\cal D}_{(0)}^{ij}(x)+g \ {\cal
D}_{(1)}^{ij} (x) + g^2 \  {\cal D}_{(2)}^{ij} (x) + \dots \, ,
\label{Dtilde}\end{equation} where ${\cal D}_{(0)}^{ij}$ denotes
the terms appearing in first line of eq.~(\ref{defD_3}). Note that
${\cal D}_{(1)}^{ij}$ contains two fermion ``impurities'', \ie\
${\cal D}_{(1)}^{ij}\approx \tau^{(i}_{AB} {\rm
Tr}(\varphi^{j)}[\lambda^A, \lambda^B]) + h.c.$. This issue will
be discussed in connection with the generalized Konishi anomaly in
Section~\ref{skonan}. Note also that the formula~(\ref{defD_3})
does not really yield an expansion in powers of $g$, because there
is an implicit $g$-dependence in each term, ${\cal D}_{(n)}^{ij}$.
In fact supergauge invariance requires the introduction of
$g$-dependent super-Wilson lines between each pair of split
points. Hence eq.~(\ref{Dtilde}) should be rather considered as a
partial operator mixing resolution which ensures that the operator
$\widetilde {\cal D}^{ij}$ does not mix with operators of naive
scale dimension $\Delta_0$ less than 4.

This situation is not peculiar to the case of the operator ${\cal D}^{ij}$.
The very same problem occurs for all composite operators where
self-contractions are not forbidden by symmetries. Indeed, one has to
define composite operators so that they do not mix with gauge invariant
operators with lower naive scale dimension. A sufficient condition to
achieve this goal for a generic purely scalar operator ${\cal O}(x)$
of naive scale dimension $\Delta_0^{({\cal O})}$ is that all $n+1$-point
correlation functions of ${\cal O}$ inserted with $n$ the fundamental
fields at non-coincident arguments
\begin{equation}
\langle \widetilde  {\cal O} (x) \ \varphi^{k_1}(y_1) \
\varphi^{k_2}(y_2) \dots \varphi^{k_n}(y_n) ) \rangle  \ =  \ 0  \
\label{O_tilde}
\end{equation}
vanish for all $n < \Delta_0^{({\cal O})}$. Let us also stress
that from the point of view of this discussion 1/2 and (would-be)
1/4 BPS operators are exceptional, since self-contractions are
absent. The former correspond to chiral primary operators (CPO's)
of the type $Tr(Z^\ell)$, with $Z=(\varphi^3 + i
\varphi^6)/\sqrt{2}$ and the latter to ``pure scalar'' operators
of the type $Tr(X^{\ell+k} Y^k)$, with $X=(\varphi^1 + i
\varphi^4)/\sqrt{2}$ and $Y=(\varphi^2 + i \varphi^5)/\sqrt{2}$.

A natural question arises from the previous considerations. Does
this modification change the results for the anomalous dimensions
and the operator mixing coefficients present in the literature?
The answer is that as far as only the order $g^2$ corrections to
the anomalous dimensions of the operators are extracted from a
perturbative calculation of two-point functions, terms beyond
${\cal D}_{(0)}^{ij}$ can be neglected in eq.~(\ref{Dtilde}). The
reason is that we can write the order $g^2$ correction to the
two-point function in the form (for illustrative purposes we shall
again refer to ${\cal D}^{ij}$, but precisely the same argument
holds for any operator)
\begin{eqnarray}
\hspace{-.8cm}&&
\langle \widetilde  {\cal D}^{ij}(x) \ \widetilde  {\cal D}^{ij}(y)
\rangle \vert_{g^2} \ =  \
 \langle {\cal D}_{(0)}^{ij}(x) \ {\cal D}_{(0)}^{ij}(y) \rangle \vert_{g^2}
+ g\langle {\cal D}_{(0)}^{ij}(x) \ {\cal D}_{(1)}^{ij}(y) \rangle
\vert_{g} + g\langle {\cal D}_{(1)}^{ij}(x) \ {\cal
D}_{(0)}^{ij}(y) \rangle \vert_{g}+ \nonumber \\\hspace{-.8cm}&&
+\ g^2\langle {\cal D}_{(0)}^{ij}(x) \ {\cal D}_{(2)}^{ij}(y)
\rangle \vert_{0} + g^2\langle {\cal D}_{(2)}^{ij}(x) \ {\cal
D}_{(0)}^{ij}(y) \rangle \vert_{0} +g^2\langle{\cal
D}_{(1)}^{ij}(x)\ {\cal D}_{(1)}^{ij}(y)\rangle\vert_{0}\, ,
\label{DDg2}\end{eqnarray} In this expansion only the first term
in the r.h.s.\ can contain divergent logarithmic terms and thus
can contribute to the anomalous dimension. Hence all order $g^2$
calculations of anomalous dimensions performed so far remain
unchanged. This however is not the case for the finite parts of
the two-point correlation functions nor for higher order
computations.

As a second example of this kind of problems let us consider, in
fact, the order $g^2$ correction to the three-point function
$\langle \widetilde{\cal D}^{ij}(x_1){\cal Q}^{kl}(x_2) {\cal
Q}^{mn}(x_3)\rangle$. This time the naive contribution
$\langle{\cal D}_{(0)}^{ij}(x_1){\cal Q}^{kl}(x_2){\cal
Q}^{mn}(x_3) \rangle \vert_{g^2}$ with the insertion of ${\cal
D}_{(0)}^{ij}$
 is non-vanishing and violates conformal invariance,
while the complete expression, where also ${\cal D}_{(1)}^{ij}$
contributes, is zero
\begin{equation}
\langle \widetilde{\cal D}^{ij}(x_1)\ {\cal Q}^{kl}(x_2)\ {\cal Q}^{mn}(x_3)
\rangle \vert_{g^2} \ = \ 0 \, ,
\label{DQQ}
\end{equation}
as expected \cite{Arutyunov:2000ku, Arutyunov:2001mh, Arutyunov:2000im, Eden:2000bk, Eden:2001ec}.
What is crucial for the present investigation is that at higher orders
in perturbation theory the difference between $\widetilde{\cal D}^{ij}$
and ${\cal D}_{(0)}^{ij}$ affects also the divergent logarithmic parts,
thus contributing corrections to the anomalous dimension.
As an illustration, let us consider the next order counter part of
eq.~(\ref{DDg2}). Again we write formulae for the particular case of the
operator ${\cal D}^{ij}$, but we remind that they are valid in general.
We get to order $g^4$
\begin{eqnarray}
\hspace{-.9cm}&&\langle \widetilde{\cal D}^{ij}(x)
\widetilde{\cal D}^{ij}(y) \rangle \vert_{g^4} =
\langle {\cal D}_{(0)}^{ij}(x) {\cal D}_{(0)}^{ij}(y) \rangle \vert_{g^4}
+ g\langle {\cal D}_{(0)}^{ij}(x) {\cal D}_{(1)}^{ij}(y) \rangle \vert_{g^3}
+g\langle {\cal D}_{(1)}^{ij}(x) {\cal D}_{(0)}^{ij}(y) \rangle \vert_{g^3}+
\nonumber\\\hspace{-.9cm}&&
+\ g^2\langle {\cal D}_{(0)}^{ij}(x) {\cal D}_{(2)}^{ij}(y)\rangle\vert_{g^2}
+ g^2\langle {\cal D}_{(2)}^{ij}(x) {\cal D}_{(0)}^{ij}(y)\rangle\vert_{g^2}
+ g^2\langle {\cal D}_{(1)}^{ij}(x)
{\cal D}_{(1)}^{ij}(y) \rangle\vert_{g^2}+ \dots \,  ,\label{DDg4}
\end{eqnarray}
where the dots stand for tree-level or O($g$) contributions which
do not produce divergent logarithmic behaviours. The
${\log}^2(\epsilon)$ divergent terms can only come from the first
term in the r.h.s. of the expansion, while contributions
proportional to ${\log}(\epsilon)$ generally come from all the
terms~\footnote{Actually in the present instance, since ${\cal
D}^{ij}$ has vanishing anomalous dimension, the fourth and the
fifth term of the expansion vanish separately, while the sum of
all the others is zero.} in (\ref{DDg4}).

One final remark concerns the conditions for an operator to be a
conformal and superconformal primary. One may wonder whether the
presence of derivative terms like $\Box{\cal Q}^{ij}$ in
eq.~(\ref{defD_3}) might spoil this property. The point is that in
the regularized theory the naive operator ${\cal D}_{(0)}^{ij}$ is
not anymore primary. It is precisely the presence of derivative
terms in the regularized (point-split) operator that makes it
primary. The same is true for the appearance of the operator
$\tau^{(i}_{AB}{\rm Tr}(\varphi^{j)}[\lambda^A, \lambda^B]) +
h.c.$ in the ${\cal D}_{(1)}^{ij}$ contribution. Only the complete
operator $\widetilde {\cal D}^{ij}$ has the correct conformal
properties as demonstrated, for example, by the computation of the
three-point function~(\ref{DQQ}).

\section{Generalized Konishi anomaly and mixing with \\ fermions}
\label{skonan}

The analysis presented in the previous section illustrates that
mixing with fermions and operators of different ``length'' (number
of constituents) is possible and in fact required beyond
one-loop~\cite{Beisert:2003ys}. Explicit resolution of this kind
of mixing leads to daunting
computations~\cite{Eden:2003sj,Beisert:2003ys}, but at least in
the planar limit and for certain classes of operators one can rely
on anomaly arguments to determine the correct mixing coefficients
to lowest non trivial order in $g$. This is our aim in this
section. The outcome of the analysis of the generalized Konishi
anomaly is a rather compact form for the mixing of certain
operators with fermion and boson ``impurities''~\footnote{The
nomenclature is taken from the BMN limit but it is valid in
general~\cite{Beisert:2002tn}.}.

In any ${\cal N}=1$ supersymmetric gauge theory with vector
multiplets $V$ in the adjoint representation and chiral multiplets
$\Phi^I$ in some representation ${\bf r}$ of the gauge group, the
Konishi supermultiplets \be {\cal K}_I{}^J = {\rm Tr}_{\bf
r}(\bar\Phi_I e^{2gV} \Phi^J ) \label{KMUL}\ee are real vector
multiplets that contain flavour currents in their
$\theta\bar\theta$ component. Naively the superfield equations of
motion yield \be {1\over 4} \bar{D}^2 {\cal K}_I{}^J = {\rm
Tr}_{\bf r}\left({\de {\cal W} \over \de \Phi^I}\Phi^J\right) \, ,
\label{konan}\ee where ${\cal W}$ is the superpotential. Together
with its hermitian conjugate, (\ref{konan}) implies \be {1\over
16}[\bar{D}^2, D^2]{\cal K}_I{}^J = {1\over 4}\bar{D}^2{\rm
Tr}_{\bf r}\left({\de \bar{{\cal W}} \over \de
\bar{\Phi}_J}\bar\Phi_I\right) -{1\over 4}D^2{\rm Tr}_{\bf
r}\left({\de {\cal W} \over \de \Phi^I}\Phi^J\right)\, . \ee While
kinetic terms are all chirally invariant, this equation expresses
the non-invariance of the interactions governed by ${\cal W}$
under chiral flavour symmetry transformations even at the
classical level.

At the quantum level the singlet current is plagued by the chiral
anomaly that is part of the ``superglueball'' multiplet ${\cal
S}$. In superfield notation the divergence equation of the singlet
current reads \be {1\over 4} \bar{D}^2 {\cal K}_I{}^J = {\rm
Tr}_{\bf r} \left({\de {\cal W} \over \de \Phi^I}\Phi^J\right) +
\delta_I{}^J{g^2  \over 16\pi^2} {\rm Tr}_{\bf r} (W^\alpha
W_\alpha)\, , \label{konan1} \ee where $W_\alpha={1\over 4}
\bar{D}^2 e^{-2gV} D_\alpha e^{2gV}$ is the chiral superfield
strength. Assuming the validity of the Adler-Bardeen theorem, one
concludes that the anomaly multiplet \be {\cal S} = {g^2 \over
16\pi^2} {\rm Tr}_{\bf r} (W^\alpha W_\alpha) \label{SSF}\ee
should not be affected by renormalization effects, \ie\ it should
yield finite operator insertions. Since ${\rm Tr}_{\bf r}(W_\alpha
W^\alpha)$ by itself is not finite, one can deduce the running
properties of the gauge coupling from the renormalization of ${\rm
Tr}_{\bf r}(W_\alpha W^\alpha)$~\cite{Amati:ft}. Although this is
a consistent scenario in ${\cal N}=1$ theories, it leads to
a contradiction in ${\cal N}=4$ SYM. Indeed, it is known that the
anomalous divergences of the currents in the ${\cal N}=4$ Konishi
multiplet ${\cal K}$, being proportional to their common anomalous
dimension (as dictated by superconformal invariance), receive
contributions not only at one-loop but also at
two-loops~\cite{Bianchi:1999ge,Bianchi:2001cm,Bianchi:2000hn,Eden:1998hh,Eden:1999kh,Eden:2000qp},
and higher orders in perturbation theory (though possibly not from
instanton
effects~\cite{Bianchi:1998nk,Bianchi:1998xk,Bianchi:2000vh,Kovacs:2003rt}).
These corrections, however, cannot be reabsorbed in the
renormalization of the coupling constant,
because the $\beta$-function of this exactly
conformal theory vanishes.

The ${\cal N}=4$ generalization of~(\ref{konan1})
is~\cite{Heslop:2001dr,Heslop:2003xu} \be {1\over 4} \bar{D}^A
\bar{D}^B {\cal K} = g {\rm Tr}([W^{AE},W^{BF}] \bar{W}_{EF}) +
{g^2  \over 16\pi^2} D_E D_F {\rm Tr}(W^{AE}W^{BF}) + {\rm O}(g^3)
\, , \label{konan4} \ee where $W^{AB}=\bar\tau^{AB}_i W^i$ is the
twisted chiral ${\cal N}=4$ SYM multiplet that starts with
$\varphi^{AB}=\bar\tau^{AB}_i \varphi^i$. Order $g^3$ 
corrections and higher are expected in~(\ref{konan4}), since ${\cal E}^{AB}
= D_E D_F {\rm Tr}(W^{AE}W^{BF})$ is a protected operator, being a
superdescendant at level two of ${\cal Q}_{{\bf 20}^\prime}$ in
the ${\cal N}=4$ supercurrent multiplet. Analogously the
tree-level term \be {\cal K}^{AB} = {\rm Tr}([W^{AE},W^{BF}]
\bar{W}_{EF})\, , \ee is the lowest component of a short 1/8 BPS
supermultiplet in free theory, but satisfies \be {1\over 4}
\bar{D}^E \bar{D}^F{\cal K}^{AB} = g \epsilon_{CDHG} {\rm
Tr}([W^{AC},W^{BD}][W^{EG},W^{FH}]) + {\rm O}(g^3) \, , \ee when
interactions are turned on. In free theory the first term in  the
r.h.s.\ is a 1/4 BPS short multiplet whose lowest component is a
``pure scalar'' operator ${\rm Tr}([X,Y]^2)$ of naive dimension
$\Delta_0=4$ belonging to the representation ${\bf 84}$ of
$SU(4)$~\cite{Bianchi:1999ge}.

Recently~\cite{Cachazo:2002ry} the Konishi anomaly equation has
been generalized to encompass the case of an ${\cal N}=1$ gauge
theory with a chiral multiplet in the adjoint representation of
the gauge group, $\Phi= \Phi^a T_a$, where $T_a$ are the
generators of the gauge group in the fundamental representation.
In this situation one can envisage the possibility of constructing
supergauge invariant generalized Konishi multiplets according to
the formula \be {\cal K}_{(\bar{n},n)} = {\rm Tr}(e^{-2gV}
\bar\Phi^{\bar{n}}e^{2gV} \Phi^n) \ , \ee with the ``standard"
Konishi multiplet being the ${\cal K}_{(1,1)}$ instance of the
above series. The authors of~\cite{Cachazo:2002ry} have mostly, if
not exclusively, concentrated their attention on the
supermultiplets ${\cal K}_{(1,n)}$ that are in a sense ${\cal
N}=1$ relatives of the BMN multiplets with two
impurities~\cite{Beisert:2002tn}. The generalized super-anomaly
equation for ${\cal K}_{(1,n)}$ reads \be {1\over 4}\bar{D}^2
{\cal K}_{(1,n)}={\rm Tr}\left({\de {\cal
W}\over\de\Phi}\Phi^n\right)+ {g^2  \over 16\pi^2} {\rm
Tr}([W_\alpha, T^a][T_a, W^\alpha] \Phi^{n-1}) + {\rm O}(g^3) \, .
\label{konan1n}\ee

In its ${\cal N}=1$ decomposition, ${\cal N}=4$ SYM comprises
three chiral multiplets $\Phi^I$, so one can consider gauge
invariant operators of the form \be {\cal
K}^{I_1...I_{n-1}}_{(1,n)} = {\rm Tr}(e^{-2gV} \bar\Phi_K e^{2gV}
\Phi^K \Phi^{I_1}\ldots \Phi^{I_{n-1}}) \, , \label{KI}\ee that
are chiral supermultiplets belonging to higher $SU(4)$
``harmonics''. In the AdS/CFT
correspondence~\cite{Maldacena:1997re,Aharony:1999ti,D'Hoker:2002aw,Bianchi:2000vh}
${\cal K}^{I_1...I_{n-1}}_{(1,n)}$ should correspond to K-K
excitations of the Konishi operator and belong  to semi-short
multiplets in the free theory that become long when interactions
are turned
on~\cite{Bianchi:2003wx,Beisert:2003te,Andrianopoli:1998ut,Dolan:2002zh}.
They satisfy a generalized anomaly equation which, in ${\cal N}=1$
notation, looks almost identical to~(\ref{konan1n}), namely \ba &&
{1\over 4}\bar{D}^2 {\cal K}^{I_1...I_{n-1}}_{(1,n)} = {\rm
Tr}\left({\de {\cal W}\over\de\Phi^I}\Phi^I \Phi^{I_1}
\ldots\Phi^{I_{n-1}}\right)  \nonumber \\ && + {3 g^2\over
16\pi^2} {\rm Tr}([W_\alpha, T^a][T_a, W^\alpha] \Phi^{I_1}\ldots
\Phi^{I_{n-1}}) + {\rm{O}}(g^3) \, . \label{konan4n} \ea

It is amusing to observe that the factor of 3 in~(\ref{konan4n})
(which is the number of chiral supermultiplets in ${\cal N} =4 $
SYM) determines the one-loop anomalous dimension of the singlet
current in the ${\cal N}=4$ Konishi multiplet and, as a
consequence of superconformal invariance, of the full multiplet.
Another consequence of the generalized anomaly is the mixing of
operators that are multi-linear in the bosons with operators
containing fermion ``impurities''. In the simplest case of the
${\cal N}=4$ Konishi multiplet this phenomenon was suggested in
ref.~\cite{Intriligator:1998ig,Intriligator:1999ff} as a way to
protect the $U(1)_B$ bonus symmetry of two-point functions and
confirmed by explicit computations in~\cite{Bianchi:2001cm} where
the relation of this phenomenon to the ${\cal N}=4$ extension of
the Konishi anomaly was stressed.

Here we see that a similar mechanism is taking place for the
higher harmonics of the Konishi multiplet that correspond to BMN
operators with two impurities~\cite{Beisert:2002tn}. In
particular, it is easy to see that only purely scalar operators in
the singlet and antisymmetric tensor representation of the $SO(4)$
subgroup of $SU(4)$ commuting with $U(1)_J$ can mix with operators
with fermion ``impurities''. Indeed the four gauginos decompose
under $SU(2) \times SU(2) \times U(1)_J$ according to \be
\lambda^A_\alpha\to\{\psi^r_{\alpha(J=+1/2)},
\chi^{\dot{r}}_{\alpha(J=-1/2)} \} \label{LAPA}\ee with
$r,\dot{r}=1,2$ and similarly for the hermitian conjugate fields.
Thus the only possibilities to build scalar operators with $\Delta
- J = 2$ with fermion ``impurities'' are given by
\begin{equation}
{\cal F}^{rs}_{(J|p)} = {\rm
Tr}(\psi^{\alpha r}_{(+1/2)} Z^p \psi^s_{\alpha(+1/2)} Z^{J-p-1})
\quad , \quad \bar{\cal F}^{\dot{r}\dot{s}}_{(J|p)} = {\rm
Tr}(\bar\chi^{\dot{r}}_{\dot\alpha(+1/2)} Z^p
\bar\chi^{\dot\alpha\dot{s}}_{(+1/2)}Z^{J-p-1})\, .
\end{equation}
${\cal F}^{[rs]}$ and $\bar{\cal F}^{[\dot{r}\dot{s}]}$ transform
as singlets, while ${\cal F}^{(rs)}$ and $\bar{\cal
F}^{(\dot{r}\dot{s})}$ transform as ${\bf 3}_L$, and ${\bf 3}_R$
of $SO(4)$, respectively. In particular it is not possible to
construct a traceless symmetric tensor (\ie\ the $({\bf 3}_L, {\bf 3}_R)$ of $SO(4)$).

For the antisymmetric tensors, which are superdescendant at level
two of the above singlets~\cite{Beisert:2002tn,Eden:2003sj}, we
thus expect
\begin{eqnarray}
{\cal O}^{[ab]+}_{(J|p)} &=& {\cal B}^{[ab]+}_{(J|p)} + c_{(J|p)} {g \over
16 \pi^2} \sigma^{[ab]}_{rs} {\cal F}^{(rs)}_{(J|p)} \nonumber\\
{\cal O}^{[ab]-}_{(J|p)} &=& {\cal B}^{[ab]-}_{(J|p)} + c_{(J|p)} {g \over
16 \pi^2} \sigma^{[ab]}_{\dot{r}\dot{s}}
\bar{\cal F}^{(\dot{r}\dot{s})}_{(J|p)}\nonumber \label{AST}
\end{eqnarray}
where
${\cal B}^{[ab]}_{(J|p)}= {\rm Tr}(\varphi^{[a} Z^p\varphi^{b]} Z^{J-p})$
with $a,b=1,\dots,4$ and $+$($-$) denotes projection onto the
(anti)self-dual (${\bf 3}_{L(R)}$) component.

 The
anomaly argument only determines the form of the superdescendants
once the superprimary are known and amounts to the addition of an
extra anomalous term to the naive second order variation 
under the ``dynamical supercharges'', \ie\ the ones
commuting with $\Delta - J$ and annihilating $Z$. The situation
for the $SO(4)$ singlets, which are  superprimary, \ie\ lowest
components, of two-impurity BMN
multiplets~\cite{Beisert:2002tn,Eden:2003sj}, is different. For
these operators  one has to rely on the methods of
Section~\ref{composite}, where the case $J=2$ corresponding to
the operator ${\cal D}_{{\bf 20}^\prime}$ has been reanalyzed. At
any rate, our present analysis suggests that the two-impurity
``dilatation operator'' of~\cite{Beisert:2003tq} should contain
terms with odd powers of $g$ even in the BMN limit, except in
those sectors, such as the $({\bf 3}_L, {\bf 3}_R)$ of $SO(4)$,
where mixing with fermion impurities is forbidden on symmetry
grounds.

Similar arguments should help resolving the mixing for the
operators described by the $SU(2|3)$ super spin chain
of~\cite{Beisert:2003ys}. The simplest instance is the well
studied case of the mixing between ${\rm Tr}([X,Y]Z)$ and ${\rm
Tr}(\lambda \lambda)$ that is resolved in terms of the operators
${\cal E}_{10}$, with vanishing anomalous dimension, and
${\cal K}_{10}$, belonging to the Konishi multiplet~\cite{Bianchi:2001cm}.
The study can be easily generalized to the mixing between putative
1/8 BPS operators of the form ${\rm Tr}(Z^l X^k Y)$ that are
either primary and thus protected or superdescendants of operators
of the form ${\rm Tr}(Z^{l-1} X^{k-1}Y\bar{Y})$ and thus mix with
${\rm Tr}(Z^{l-1} X^{k-1}\lambda \lambda)$. An analysis of the
mixing in this sector  should help clarifying some of the issues
left open in~\cite{Beisert:2003ys}.

\section{The operator mixing resolution}
\label{opemixres}

Let the operators $\widetilde {\cal  O}_p(x,\epsilon)$, with
$p=1,\ldots,n$, be a basis of bare point-split regularized
operators of naive dimension $\Delta_0$, which are properly
defined as discussed in Section~\ref{composite}. We would like to
diagonalize the matrix of their two-point functions and find the
corresponding anomalous dimensions. Assume that we have computed
the two-point functions to some order in perturbation
theory~\footnote{To this purpose, we can safely neglect any
possible dependence on the vacuum angle $\vartheta$.}. The result
of the calculation has the form
\begin{equation}
\langle \widetilde {\cal O}_p(x,\epsilon) \
\widetilde{\cal O}_q^{\dagger}(y,\epsilon)\rangle  =  f_{pq}
\left({\epsilon^2\over (x-y)^2 },g\right) \ {1\over [(x-y)^2]^{\Delta_0}}\, ,
\label{2ptbare}
\end{equation}
where $f_{pq}$ is an hermitian matrix depending on the operator basis we
have chosen. In fact, since complex operators come in pairs with
the same anomalous dimension we can always choose a basis in which
$f_{pq}$ is real and symmetric.

The renormalized operators $\widehat {\cal O}_p$
which have well defined anomalous dimensions $\gamma_p(g^2)$
are linear combinations of the operators $\widetilde {\cal O}_q$
\begin{equation}
\widehat {\cal O}_p(x,\mu)  = \sum_q Z_{pq}(\epsilon^2 \mu^2,g)
\ {\widetilde {\cal O}}_q(x,\epsilon)\, ,
\label{opren}
\end{equation}
where the auxiliary scale $ \mu $ is an artifact of the
perturbative expansion and plays the role of subtraction point.
Its presence contradicts neither scale nor conformal
invariance~\cite{Bianchi:1999ge}. As discussed in Section~\ref{composite},
we shall assume that the matrix $Z$ has an inverse. Scale
invariance completely determines the two-point functions of
$\widehat {\cal O}_p (x,\mu)$ to be
\begin{equation}
\langle\widehat{\cal O}_p(x,\mu) \ \widehat{\cal
O}_q^{\dagger}(y,\mu)\rangle  = {\delta_{pq} \over [(x-y)^2
]^{\Delta_0} [(x-y)^2 \mu^2 ]^{\gamma_p(g^2)}}\, , \label{2ptren}
\end{equation}
where we have separately indicated the dependence on the naive and the
anomalous dimension. Let us stress that, while $f_{pq}$ and $Z_{pq}$ can
in general depend on both even and odd powers of the coupling constant $g$,
the physical anomalous dimensions, $\gamma$, can only be function of $g^2$.
Compatibility among the above three equations implies (dropping indices)
\begin{equation}
Z(\epsilon^2 \mu^2,g)\ f\left({\epsilon^2\mu^2\over (x-y)^2\mu^2
},g\right)\ Z^{\dagger}(\epsilon^2 \mu^2,g)  = \left[(x-y)^2 \mu^2
\right]^{-\Gamma(g^2)} \, ,\label{renorm1}
\end{equation}
where $\Gamma(g^2)$ is the diagonal matrix of anomalous
dimensions. For future convenience we introduced a  $\mu$
dependence in both the numerator and the denominator of the
argument of $f$. Since there exists a basis in which both $f$ and
$Z$ are real, the (diagonal) elements of $\Gamma(g^2)$, which
represent the sought for anomalous dimensions, are also all real.

It is useful to compute eq.~(\ref{renorm1}) at two special points, namely

\noindent 1) $\epsilon^2 \mu^2=1$ and $(x-y)^2 \mu^2 = 1 / u$ which yields
\begin{equation}
Z(1,g)  \  f(u,g) \ Z^{\dagger}(1,g)  =
u^{\Gamma(g^2)} \, ,
\label{renorm2}
\end{equation}
\noindent 2) $\epsilon^2 \mu^2=u$ and $(x-y)^2 \mu^2 = 1 $ which yields
\begin{equation}
Z(u,g) \   f(u,g) \  Z^{\dagger}(u,g)  =  1 \ .
\label{renorm3}
\end{equation}
Note that these two equations have to be simultaneously fulfilled.
Their consistency implies that the function $f(u,g)$ has to satisfy
(for any choice of basis for the set of regularized operators
$\widetilde {\cal O}^i$ such that the matrix $Z$ has an inverse)
\begin{equation}
  f(u,g)   =  f(1,g) \  f^{-1} \left({1 \over u},g \right) \ f(1,g) \, ,
\label{consistentf}
\end{equation}
where $f^{-1}$ is the inverse of the matrix $f$. Assume that we have found
a solution, $Z(1,g)$, of eq.~(\ref{renorm2}). Then it is immediate to see that
\begin{equation}
  Z(u,g)   =  u^{-{1 \over 2} \, \Gamma(g^2)} \  Z(1,g)
\label{zl}
\end{equation}
solves eq.~(\ref{renorm3}). The last relation has a simple intuitive
explanation, one first defines by means of $Z(1,g)$ operators with well
defined scale dimension, then the field-theoretical renormalization step
amounts to a simple rescaling by the factor
$(\epsilon^2 \mu^2)^{-{1 \over 2}\Gamma(g^2)}$. Since in general (for
non-degenerate $\Gamma(g^2)$) the solution for $Z$ is unique, it will be given
by eq.~(\ref{zl}). Hence the $u$ dependence in $Z(u,g)$ factorizes and we have
to solve only eq.~(\ref{renorm2}) for the unknown $Z(1,g)$ and $\Gamma(g^2)$
once the function $f(u,g)$ is known.

\subsection{Order by order analysis}
\label{ordbyord}

In perturbation theory $f(u,g)$, $Z(u,g)$ and $ \Gamma(g^2)$ admit
an expansion in powers of the coupling constant. Introducing for
short the definition $\ell={\rm log}(u)$, we get the obvious
expansions
\begin{eqnarray}
\hspace{-1.2cm}&&f(u,g)   =  f_{00} +  g f_{10}+ g^2(f_{20}+\ell f_{21})
+ g^3(f_{30}+\ell f_{31}) +g^4(f_{40}+\ell f_{41}+\ell^2 f_{42})+
\dots  \, ,\label{fexp}\\
\hspace{-1.2cm}&&Z(1,g)   =  Z_0 + g Z_1 +g^2 Z_2+g^3 Z_3+g^4 Z_4+\dots  \, ,\\
\label{zexp}
\hspace{-1.2cm}&&\Gamma(g^2)   =  g^2 \Gamma_1 +g^4 \Gamma_2+\dots  \, .
\label{Gexp}
\end{eqnarray}
Let us substitute these expressions in eq.~(\ref{renorm2}) and consider
for the moment only tree-level terms and the terms proportional to
$g^2 \cdot\ell$. We get the equations
\begin{eqnarray}
 {\rm tree:} &&\qquad Z_0 \ f_{00} \ Z_0^{\dagger}   =  1  \nonumber \\
 {g^2 \cdot \ell: }&& \qquad Z_0  \ f_{21} \ Z_0^{\dagger}   =  \Gamma_1
\label{trg2}
\end{eqnarray}
which determine $Z_0$ and $\Gamma_1$. In fact we have two matrix
equations involving hermitian $n \times n $ matrices. Hence they
lead to $n(n+1)$ scalar equations for $n^2$ [$Z_0$] + $n$
[$\Gamma_1$] unknown. In the next subsection we shall obtain the
general solution of the system~(\ref{trg2}). In order to go to
higher order it is convenient to make a change of operator basis
which significantly simplifies the formulae. Let us rotate the
original basis, by a solution of eqs.~(\ref{trg2}). In this way
tree-level and $g^2 \cdot\ell$ terms will be diagonal and we get
in the new basis
\begin{equation}
Z_0 = 1 \quad , \qquad f_{00} = 1 \quad , \qquad f_{21} = \Gamma_1\, .
\end{equation}
The remaining conditions coming from terms up to order $g^4$ now give the
following set of relations
\begin{eqnarray}
{g \  : }& \quad
f_{10} + Z_1  + Z_1^{\dagger} =  0  \qquad &\rightarrow  Z_1^H \\
{g^3 \cdot\ell  : }& \quad
f_{31} + Z_1 \Gamma_1 + \Gamma_1 Z_1^{\dagger} = 0\qquad &\rightarrow Z_1^A \\
{g^2  \ : }& \quad
 f_{20} + Z_1 f_{10} + f_{10} Z_1^{\dagger} + Z_1 Z_1^{\dagger}
+Z_2+Z_2^{\dagger} =  0 \qquad &\rightarrow  Z_2^H \\
{g^4 \cdot  \ell : }&  \quad
 f_{41} + Z_1 f_{31} + f_{31} Z_1^{\dagger} + Z_2 \Gamma_1
+ \Gamma_1 Z_2^{\dagger}+ Z_1 \Gamma_1 Z_1^{\dagger} =
\Gamma_2  \qquad &\rightarrow Z_2^A, \Gamma_2\\{g^4 \cdot  \ell^2 : }& \quad
f_{42}  =  {1 \over 2}  (\Gamma_1)^2 & \label{g2g4}
\end{eqnarray}
The first pair of equations determine the hermitian, $Z_1^H$, and
the anti-hermitian, $Z_1^A$, part of $Z_1$, respectively, as
indicated by the symbols after the arrow. The second pair of
equations determine the hermitian, $Z_2^H$, and the
anti-hermitian, $Z_2^A$, part of $Z_2$, as well as the order $g^4$
correction to the anomalous dimensions, $\Gamma_2$. The counting
of equations {\it vs} parameters is as before: we have $n(n+1)$
equations for $n^2$ [$Z_2$] + $n$ [$\Gamma_2$] unknown. The last
equation is a consistency condition and is automatically satisfied
if $f$ satisfies~(\ref{consistentf}).

The pattern repeats itself to every  order. The constant term at
order $g^{n}$ and the term with coefficient $g^{n+2}\cdot\ell$
determine $Z_n$ and (for even $n$) also $\Gamma_{n}$. All the
terms containing  higher powers of $\ell$ give just consistency
conditions which are trivially satisfied if~(\ref{consistentf}) is
true. Let us stress that once $Z_0$ is found, all the remaining
equations are linear. Thus in general the only residual freedom is
an arbitrary rotation in the subspaces of operators with the same
(anomalous) dimension, if there are such degenerate operators that
cannot be discriminated by other ``good'' quantum numbers. Actually
we will see in sect.~\ref{g2comput} that this phenomenon takes
place.

\subsection{Solving tree-level mixing}
\label{treeres}
In this subsection we shall give the general solution of the tree-level
mixing problem defined by eqs.~(\ref{trg2}) for the unknown $Z_0$ and
$\Gamma_1$, where $f_{00}$ and $f_{21}$ are given hermitian matrices.

Let $E_0$ be the matrix of the (orthonormal) eigenvectors of the hermitian
matrix $f_{00}$. By definition $E_0$ diagonalizes $f_{00}$.
\begin{equation}
 E_0 \ f_{00} \ E_0^{\dagger}   =  h_0  \, ,
\label{E0}
\end{equation}
where $h_0$ is a diagonal matrix. Since $f_{00}$ is the matrix of the
tree-level two-point functions, unitarity implies that all the diagonal
entries of $h_0$ are strictly positive. Hence $h_0$ is invertible and we
can unambiguously define the matrix $(h_0)^{-{1 \over 2}}$ by taking \eg\
its positive square root. The formulae
\begin{eqnarray}
(h_0)^{-{1 \over 2}}E_0\  f_{00}\ E_0^{\dagger}(h_0)^{-{1\over 2}}&=&1 \, ,
\nonumber \\
(h_0)^{-{1\over 2}}E_0\ f_{21} \ E_0^{\dagger}(h_0)^{-{1\over 2}}&=&F_{21}\, ,
\label{D0E0}\end{eqnarray}
where $F_{21}$ is again a hermitian matrix, then follow.

Let now $E_1$ be the matrix of the (orthonormal) eigenvectors of
the hermitian matrix $F_{21}$. By definition $E_1$ diagonalizes
$F_{21}$, so sandwiching the system (\ref{D0E0}) between $E_1$ and
$E_1^{\dagger}$ we obtain
\begin{eqnarray}
E_1 (h_0)^{-{1 \over 2}}E_0\ f_{00}\ E_0^{\dagger}(h_0)^{-{1 \over 2}}
E_1^{\dagger} &=& E_1 E_1^{\dagger} \ = \  1 \, ,
\nonumber \\
E_1 (h_0)^{-{1 \over 2}} E_0 \  f_{21} \ E_0^{\dagger}(h_0)^{-{1\over 2}}
E_1^{\dagger}  &=& E_1 \ F_{21}  \ E_1^{\dagger} \   =  \ \Gamma_1 \, .
\label{E1D0E0}\end{eqnarray}
Putting together the previous results, we conclude that the matrix
\begin{equation}
Z_0 \ = \ E_1 \ (h_0)^{-{1 \over 2}} \ E_0
\label{Z0}
\end{equation}
diagonalizes simultaneously $f_{00}$ and $f_{21}$, hence it is the solution
to the tree-level mixing problem posed by eqs.~(\ref{trg2}). The order $g^2$
correction to the anomalous dimension, $\Gamma_1$, can then be read off
from the second of eqs.~(\ref{E1D0E0}).

Let us note that, if one is interested only in the values of the
anomalous dimensions $\Gamma_1$, they can be obtained in a much simpler
way by computing the eigenvalues of the matrix
\begin{equation}
(f_{00})^{-1} \ f_{21}\, .
\label{f0if1}
\end{equation}
We prefer to use the more complicated method described in this
section, since it gives us also the mixing matrix $Z_0$, \ie the
explicit form of the corresponding operators.

We end the section with two remarks. Note that the role of tree
level and order $g^2$ terms is not symmetric in the above
formulae. Since in general there is no positivity constraint for
the order $g^2$ contributions, $f_{21}$ and $\Gamma_1$ may have
vanishing eigenvalues and may not be invertible. It should also be
said that in most cases the tree-level mixing matrix $Z_0$ can be
computed only numerically, because the eigenvalue problem is highly
nontrivial and cannot be solved analytically.

\section{Anomalous dimensions of singlet operators with $\Delta_0=6$}
\label{anodim6}

Let us  illustrate some applications of the general method
developed in Section~\ref{opemixres}. We start with  the
computation of the anomalous dimensions of the purely scalar
$SU(4)$ singlet operators with $\Delta_0=6$. This example is
interesting for several reasons. On the one hand $\Delta_0=6$ is
the first case in which single, double and triple trace operators
come into play, so one can study the pattern of large $N_c$
suppression of the mixing between these three different types of
operators. On the other hand the number of operators is
sufficiently large (fifteen), so that one can really test the
effectiveness of the mixing resolution methods proposed in this
paper.

Last, but not least, we find a rather surprising result in this
sector of the theory, namely we find one operator which has
vanishing order $g^2$ correction to its anomalous dimension. Since
the corresponding supermultiplets is long even in the free limit
$g=0$~\footnote{We thank B. Eden and  E. Sokatchev  for
discussions on this point.}, there is no known mechanism which
should protect this operator. Although the vanishing of the
anomalous dimension of this operator may look as a one-loop
accident, we have confirmed it by a two-loop computation, \ie\ to
order $g^4$~\cite{longT}. At present we don't know of any simple
explanation for this remarkable result. It might point to some
deep and as yet uncovered property of ${\cal N}=4$ SYM. Certainly
this discovery deserves a better understanding.

For $N_c \geq 6 $ there are  $15$ distinct $SU(4)$ singlet operators
of naive conformal dimension $\Delta_0=6$ built in terms of only the
elementary scalar fields, $\varphi^i$, without derivatives.
For $N_c$ = 2, 3,
4 and 5 the number of operators of this kind is 3, 8, 13 and 14 respectively.

As explained in Section~\ref{opemixres}, for the task of computing
the order $g^2$ anomalous dimensions of these operators, we shall
have only to consider their tree-level mixing and compute to order
$g^2$  the matrix of their two-point correlation functions.
Furthermore, for the reasons discussed after eq.~(\ref{DDg2}), we
can ignore at the order we work the mixing of operators built only
in terms of elementary scalar fields with operators containing
derivatives, fermions, or the gauge field strength, $F_{\mu \nu}$.

We can choose as a basis in the $15$ dimensional space of the scalar operators
of interest the following set of operators (summation over repeated indices
goes from 1 to 6 )
\begin{eqnarray}
{\cal O}_1&=&{\rm Tr}(\varphi^i\varphi^j\varphi^k\varphi^i\varphi^j\varphi^k)
\nonumber \\
{\cal O}_2&=&{\rm Tr}(\varphi^i\varphi^j\varphi^k\varphi^j\varphi^i\varphi^k)
\nonumber \\
{\cal O}_3&=&{\rm Tr}(\varphi^i\varphi^i\varphi^j\varphi^k\varphi^j\varphi^k)
\nonumber \\
{\cal O}_4&=&{\rm Tr}(\varphi^i\varphi^i\varphi^j\varphi^k\varphi^k\varphi^j)
\nonumber \\
{\cal O}_5&=&{\rm Tr}(\varphi^i\varphi^i\varphi^j\varphi^j\varphi^k\varphi^k)
\nonumber \\
\nonumber \\
{\cal O}_6 &=& {\rm Tr}(\varphi^i  \varphi^j) \
{\rm Tr}( \varphi^i \varphi^k  \varphi^j \varphi^k) \nonumber \\
{\cal O}_7 &=& {\rm Tr}(\varphi^i  \varphi^j ) \
{\rm Tr}( \varphi^i \varphi^j  \varphi^k \varphi^k) \nonumber  \\
{\cal O}_8 &=& {\rm Tr}(\varphi^i  \varphi^i ) \
{\rm Tr}(\varphi^j \varphi^k  \varphi^j \varphi^k) \nonumber  \\
{\cal O}_9 &=& {\rm Tr}(\varphi^i  \varphi^i)  \
{\rm Tr}(\varphi^j \varphi^j  \varphi^k \varphi^k) \nonumber \\
\nonumber \\
{\cal O}_{10} &=& {\rm Tr}(\varphi^i  \varphi^j \varphi^k ) \
{\rm Tr}(\varphi^i  \varphi^j \varphi^k) \nonumber \\
{\cal O}_{11} &=& {\rm Tr}(\varphi^i  \varphi^j \varphi^k ) \
{\rm Tr}(\varphi^i  \varphi^k \varphi^j) \nonumber \\
{\cal O}_{12} &=& {\rm Tr}(\varphi^i  \varphi^i \varphi^k ) \
{\rm Tr}(\varphi^j  \varphi^j \varphi^k) \nonumber \\
\nonumber \\
{\cal O}_{13} &=& {\rm Tr}(\varphi^i  \varphi^j) \
{\rm Tr}( \varphi^i \varphi^k ) \
{\rm Tr}(  \varphi^j \varphi^k) \nonumber \\
{\cal O}_{14} &=& {\rm Tr}(\varphi^i  \varphi^i) \
{\rm Tr}( \varphi^j \varphi^k ) \
{\rm Tr}(  \varphi^j \varphi^k) \nonumber \\
{\cal O}_{15} &=& {\rm Tr}(\varphi^i  \varphi^i) \
{\rm Tr}( \varphi^j \varphi^j ) \
{\rm Tr}(  \varphi^k \varphi^k)
\label{O1_15}
\end{eqnarray}

\subsection{The order $g^2$ calculation}
\label{g2comput}

Perturbative calculations are performed in the ${\cal N} = 1$
formulation of ${\cal N} = 4$ SYM, as described in~\cite{Bianchi:2002rw}.
We  rewrite all the above operators in the ${\cal N} = 1$ language
with the help of the identity ($a,b$ are adjoint colour indices)
\begin{equation}
\sum_{i=1,\ldots,6} \varphi^i_a \varphi^i_b = \sum_{I=1,2,3}
(\phi^I_a \phi^{\dagger}_{I,b}+\phi^I_b \phi^{\dagger}_{I,a})\, .
\label{traceN1}
\end{equation}
As we do not make use of the Wess-Zumino gauge fixing, all the operators
have to be made gauge invariant by including the appropriate vector field
exponents. This amounts to the substitution
\begin{equation}
\phi^I (x) \to {\rm e}^{-g c(x)} \ \phi^I (x)  \ {\rm e}^{g c(x)} \, ,\qquad
\phi^{\dagger}_{I} (x) \to {\rm e}^{g c(x)}\
\phi^{\dagger}_{I} (x) \ {\rm e}^{-g c(x)} \, ,\label{gaugec}
\end{equation}
where $c(x)$ is the lowest component of the ${\cal N} = 1$ vector field.
We regularize the operators by point-splitting, which, as already explained,
allows us to keep also track of quadratic divergences and consistently
subtract them out.

All the order $g^2$ perturbation theory integrals entering the
calculation are proportional to the standard massless box integral
\cite{Bianchi:1999ge}, so the most complicated part of the whole
procedure is the evaluation of colour traces. Once this is done,
one obtains in the basis~(\ref{O1_15}) the explicit expression of
the two $15 \times 15$ matrices, $f_{00}$ and $f_{21}$, which are
needed in eq.~(\ref{trg2}). Using eqs.~(\ref{Z0})
and~(\ref{E1D0E0}), we resolve the tree-level mixing by
identifying the m.r. operators $\widehat{\cal O}_p$ and compute
the order $g^2$ corrections to their anomalous dimensions.
%%%%%%%%%%%%%%%%%%%%%%%
\begin{figure}[!htb]
%[!htbp]
% \begin{minipage}[t]{\linewidth}
    \centering
    \includegraphics[width=1\linewidth]{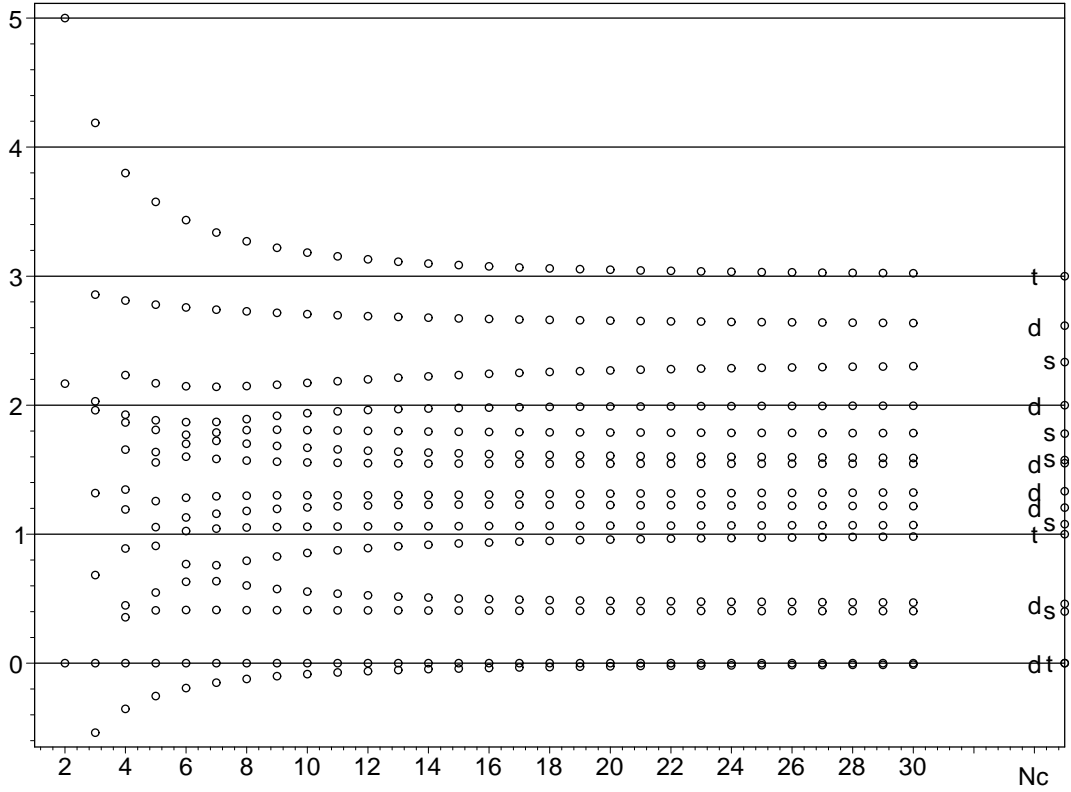}
\caption{The ratio $\gamma_1^{\widehat {\cal
O}_p}/\gamma_1^{\calK}$, $p=1,\ldots,15 $, of the $SU(4)$ singlet
operators as functions of $N_c$. $\gamma_1^{\cal K}={3 N_c\over 4
\pi^2}$ is the order $g^2$ anomalous dimension of the Konishi
multiplet.}
  \label{diagrA}
% \end{minipage} \\ [2.5cm]
\end{figure}
%%%%%%%%%%%%%%%%%%%%%%%%%%%%%%%%%%%%%%%%%%
In Figure~\ref{diagrA}  we plot the ratio of the order $g^2$
anomalous dimension of the operators $\widehat {\cal O}_p$ to the
order $g^2$ anomalous dimension of the Konishi multiplet as
function of the number of colours $N_c$ from 2 to 30. The points
at the extreme right of the figure correspond to $N_c=\infty$. The
labels $s$, $d$ and $t$ denote mostly single, double and triple
trace operators in the large $N_c$ limit. The anomalous dimensions
(in units  of ${g^2 N_c \over 4 \pi^2}$) are  the roots of the
polynomial
\begin{eqnarray}
&& \gamma \ [ 8 \gamma ^{14} - 488 \gamma ^{13} - 4 ( - 3369 + 691
\nu ^{2}) \gamma ^{12} + ( - 222905 + 136548 \nu ^{2}) \gamma
^{11}   + \nonumber \\
&&  (2461929 - 3029657 \nu ^{2} + 219960 \nu ^{4})\, \gamma ^{10}
- (8841500 \nu ^{4} + 19153227 - 39898971 \nu ^{2})
\gamma ^{9}  - \nonumber \\
&&  (347174976 \nu ^{2} - 107844797 - 157896850 \nu ^{4}
 + 7954400 \nu ^{6}) \gamma ^{8} + ( - 1656062605 \nu ^{4} - \nonumber \\
&&  444279889 + 2102167847 \nu ^{2} + 240075800 \nu ^{6}) \gamma
^{7}
+ 2 (668238125 +  \nonumber \\
&&  5674358030 \nu ^{4} - 1578288500 \nu ^{6} - 4542787849 \nu
^{2} + 78212000 \nu ^{8}) \gamma ^{6}
- (3264508000 \nu ^{8} + \nonumber \\
&& 53358746505 \nu ^{4 } - 23855595450 \nu ^{6} + 2892377256
 - 28282053929 \nu ^{2})\gamma ^{5}  -   \\
&&  15( 4208678698 \nu ^{2} - 11669730693 \nu ^{4} - 290934306
+ 98688000 \nu ^{10} - 1935494000 \nu ^{8}  + \nonumber \\
&&  7638118620  \nu ^{6})\gamma ^{4}  + 45(386368000 \nu ^{10} -
8804297250 \nu ^{4} + 8005870340 \nu ^{6} +
 \nonumber \\
&&  2201578547 \nu ^{2} - 3215135000 \nu ^{8} - 96244863)\gamma
^{3}  + 675( - 1067814772 \nu ^{
6} -  \nonumber \\
&& 154776661 \nu ^{2} + 3738354 + 649659680 \nu ^{8} + 873783513
\nu ^{4} - 132636800  \nu ^{10} + \nonumber  \\
 &&
 7680000 \nu ^{12})\gamma ^{2}  -
4050( 192054180 \nu ^{8} + 127206362 \nu ^{4} - 203160193 \nu ^{6}
 - 16503657 \nu ^{2} +   \nonumber \\
 && 3200000 \nu ^{12}
  - 66128800 \nu ^{10} + 160308)\gamma + \nonumber \\
 &&  607500 \nu ^{2} (9 - 9 \nu ^{2} + 22 \nu ^{4})\,(
 - 3577 + 32202 \nu ^{2} - 31656 \nu ^{4} + 6400 \nu ^{6})]  \quad ,
 \nonumber
\end{eqnarray}
where $\nu= 1/N_c$. The values of the anomalous dimensions for the 5 single
trace operators at $N_c=\infty$ agree with the ones obtained in
ref.~\cite{Beisert:2003tq} by diagonalizing the dilatation operator.

Two observations which are not obvious from the above picture are
the following. First of all, although some of the solutions become
rather close to each other, in particular in the range for $N_c$
from 6 to 10, they never intersect. Secondly, as soon as $N_c \geq
6 $ (to avoid accidental degeneracies) the ratio of the sum of the
order $g^2$ anomalous dimensions to the Konishi one is independent
of $N_c$, \ie\
\begin{equation}
\sum_{p=1}^{15}{\gamma_1^{\widehat{\cal O}_p} \over\gamma_1^{\cal
K}}={61 \over 3} \, . \label{sumgamma}
\end{equation}
This $N_c$ independence seems to be a rather general property. In
fact the matrix $(f_{00})^{-1} \ f_{21}$ of eq.~(\ref{f0if1}),
whose eigenvalues give the anomalous dimensions has a particularly
simple $N_c$-dependence. If we denote by $(6)$ the single trace,
by $ (2 \vert  4)$ and $(3 \vert  3)$ the two different types of
double trace and by $ (2 \vert  2 \vert  2) $ the triple trace
operators defined in eq.~(\ref{O1_15}) (in parenthesis the number
of scalar fields in each trace is indicated), then the form of the
matrix $(f_{00})^{-1} \ f_{21}$ is particularly illuminating as it
has the block structure shown below
\begin{equation}
\begin{array}{cccccc}
& & (6)  & (2 \vert   4) & (3 \vert 3)  & (2 \vert  2 \vert  2) \\
  \\
(6)    & & \# N_c & \#  & \#  & 0 \\
 \\
(2 \vert   4) & & \#   & \# N_c & 0 & \#  \\
 \\
(3 \vert  3) & & \#  & 0 & \# N_c & 0 \\
 \\
(2 \vert 2 \vert 2) & & 0  & \# & 0 & \# N_c \\
\end{array}
\label{f0if1block}
\end{equation}
where $\#$ denotes $N_c$ independent numeric entries. One
immediately concludes that the trace of this matrix (which is
equal to the sum of the (order $g^2$) anomalous dimensions) will
be proportional to $N_c$, while the ratio~(\ref{sumgamma}) will be
independent of $N_c$, as the factor $N_c$ cancels between
numerator and denominator.

This suggestive form of the matrix~(\ref{f0if1block}) has a simple
interpretation in terms of 't~Hooft double-line notation or better
in a dual string
description~\cite{Kristjansen:2002bb,Constable:2002hw,Beisert:2002bb,Constable:2002vq}.
In the latter, where operators are represented as closed strings,
the $N_c$ dependence in the matrix~(\ref{f0if1block}) is dictated
by the number of string splittings/joinings necessary to get a
connected world-sheet diagram representing the corresponding
two-point function. In fact, each closed string splitting/joining
event costs a factor  $g_s\approx 1/N_c$. Notice that some
entries which appear  as zeroes are actually suppressed by
$1/N_c^2$ factors. These corrections are, however, not visible at
the order we are working, which is $g^2 \approx g_s\approx 1/N_c$.

The most surprising feature in  Figure~\ref{diagrA} is, however,
the presence of one operator which has no order $g^2$ correction
to its anomalous dimension. It is a triple trace operator which we
shall denote by ${\cal T}(x)$. In the operator basis introduced in
eq.~(\ref{O1_15}) it takes the form
\begin{equation}
{\cal T}(x)={\cal O}_{13}-{1\over 2}{\cal O}_{14}+{1\over 18}{\cal O}_{15}\, .
\label{defT2}
\end{equation}
Two further equivalent, but much more suggestive (since they yield
${\cal T}(x)$ in terms of protected operators only) representations of
${\cal T}(x)$ are
\begin{equation}
 {\cal T}(x) \quad = \ \sum_{i,j,k=1,\ldots,6} \ : {\cal Q}^{ij}(x) \
{\cal Q}^{ik}(x) \ {\cal Q}^{jk}(x) : \quad
 = \ \sum_{i,j=1,\ldots,6} \ : {\cal Q}^{ij}(x) \ {\cal D}^{ij}(x)  : \ ,
\label{defT}
\end{equation}
where ${\cal Q}^{ij}$ are the lowest components of the
supercurrent multiplet defined in eq.~(\ref{defQ_1}), while ${\cal
D}^{ij}(x)$ is the dimension $\Delta=4$ double trace protected
operator defined in eq.~(\ref{defD_1}). We remind also that as far
as one is interested only in the order $g^2$ anomalous dimensions,
the naive prescription of the normal ordering  can be used (\ie\
$: \; : \  \equiv \ $ no self-contractions).

The vanishing of the order $g^2$  anomalous dimension of the
operator ${\cal T}(x)$ is completely unexpected, because, as we
already pointed out, it belongs, even in free theory, to a long
supermultiplet and there is no known mechanism at work which
might protect it. An exhaustive search throughout the whole set
of $\Delta_0=6$ purely scalar operators~\cite{longT} shows that
${\cal T}(x)$ is the only such operator with vanishing order $g^2$
anomalous dimension which is not protected by any known shortening
condition. This puzzling result is confirmed and made more
dramatic by the fact that one also finds a vanishing order $g^4$
correction to the anomalous dimension of ${\cal T}(x)$.

We end this section by noticing that the procedure illustrated
before can be used to resolve the order $g^2$ mixing also for
sets of operators belonging to $SU(4)$ representations other than
the singlet. As an example we wish to discuss the case of the
operators of naive conformal dimension $\Delta_0=6$ in the
representation ${\bf 20}^\prime$ of $SU(4)$. A straightforward but
tedious analysis shows that for $N_c \geq 6 $ there are 26
operators made of scalars. For $N_c$ = 2, 3, 4 and 5 the number of
operators of this kind is 4, 13, 23 and 25 respectively.
%%%%%%%%%%%%%%%%%%%%%%%
\begin{figure}[!htb]
%[!htbp]
% \begin{minipage}[t]{\linewidth}
    \centering
    \includegraphics[width=1\linewidth]{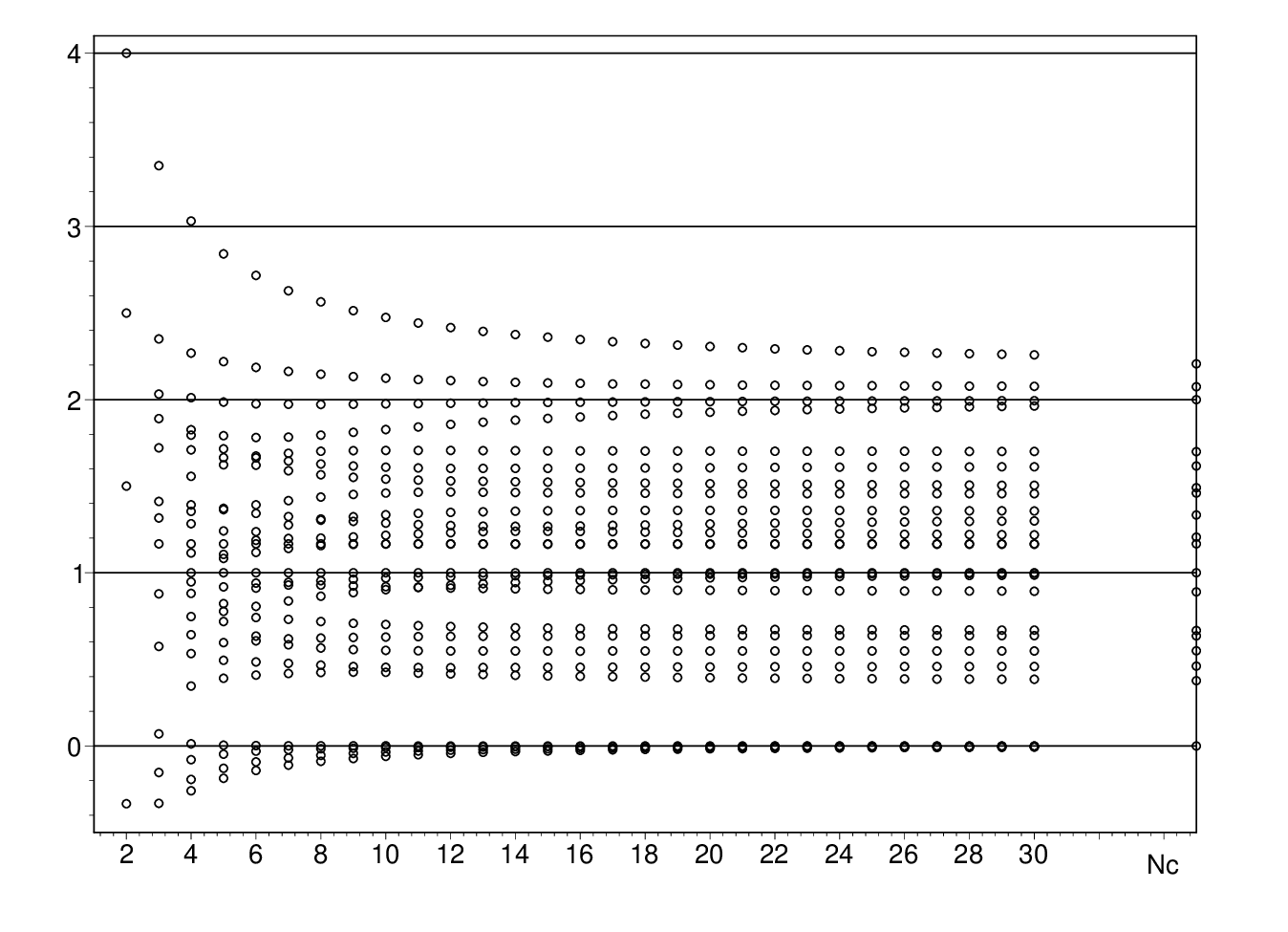}
\caption{The ratio $\gamma_1^{\widehat {\cal
O}_p}/\gamma_1^{\calK}$, $p=1,\ldots,26 $, of the operators in
${\bf 20^{\prime}}$ as functions of $N_c$. $\gamma_1^{\cal K}={3
N_c\over 4 \pi^2}$ is the order $g^2$ anomalous dimension of the
Konishi multiplet.}
  \label{diagrB}
% \end{minipage} \\ [2.5cm]
\end{figure}
%%%%%%%%%%%%%%%%%%%%%%%%%%%%%%%%%%%%%%%%%%
In Figure~\ref{diagrB} we plot the ratio of the order $g^2$
anomalous dimension of these operators to the order $g^2$
anomalous dimension of the Konishi multiplet as function of the
number of colours $N_c$ from 2 to 30. The points at the extreme
right of the figure correspond to $N_c=\infty$. As can be seen,
for finite $N_c$  all operators have non-vanishing order $g^2$
anomalous dimensions. Another surprise is  however in store for
us. In fact, we find that one operator, which in the large $N_c$
limit is dominantly double trace, hence cannot belong to the
Konishi supermultiplet, has exactly  the same order $g^2$
anomalous dimension as the Konishi supermultiplet for all values
of $N_c$. What is the origin of this degeneracy and whether it
persists at higher orders is still an open problem.

\subsection{The order $g^4$ calculation}
\label{g4comput}

In this subsection we sketch the argument which leads to the
conclusion that the order $g^4$ correction to the anomalous
dimension of the operator ${\cal T}$ vanishes. To this end let us
first note that the order $g^2$ correction to the four-point
function of ${\widetilde {\cal T}}(x)$ (the definition of the
operators ${\widetilde {\cal O}}(x)$ is as in eq.~(\ref{O_tilde})
\ie they have vanishing 2-point functions with all lower
dimensional operators) and three ${\cal Q}^{ij}$'s also vanishes,
namely
\begin{equation}
\langle {\widetilde {\cal T}}(x_1) \ {\cal Q}^{ij}(x_2) \
{\cal Q}^{ik}(x_3) \ {\cal Q}^{jk}(x_4)
\rangle \vert_{g^2} \ = \ 0 \, .
\label{TQQQ}
\end{equation}
To realize how unusual this result is we recall that in general even the
four-point functions of four protected 1/2 BPS operators are corrected
at order $g^2$.

Furthermore, one finds
\begin{equation}
\langle {\widetilde {\cal O}_{\ell}}(x_1)\
{\cal Q}^{ij}(x_2)\ {\cal Q}^{ik}(x_3) \ {\cal Q}^{jk}(x_4)
\rangle \vert_{g^2} \ = \ 0 \, ,\label{OQQQ}
\end{equation}
where ${\widetilde {\cal O}_{\ell}}(x_1)$ is an arbitrary (not
necessarily purely  scalar, like the operators in
eqs.~(\ref{O1_15})) $SU(4)$ singlet scalar operator of naive
dimension $\Delta_0=6$. This means that the operator $\widetilde
{\cal T}$ is the only $SU(4)$ singlet scalar conformal primary
operator of naive dimension $\Delta_0=6$ which appears in the OPE
of three operators ${\cal Q}^{ij}$ up to order $g^2$. The
importance of this conclusion is that we can exploit the OPE of
three  ${\cal Q}^{ij}$'s to give a rigorous definition of the
renormalized operator ${\widehat {\cal T}}(x)$ through the formula
\begin{equation}
{\widehat {\cal T}}(x) \  =  {\rm OPE} \left( \ \sum_{i,j,k} \
{\cal Q}^{ij}(x+\epsilon) \ {\cal Q}^{ik}(x) \ {\cal
Q}^{jk}(x-\epsilon) \right) {\Big \vert}_{\ \Delta_0=6} \ \  ,
\label{OPE_T}
\end{equation}
where the projection on the dimension $\Delta_0=6$ contribution in
the OPE means the subtraction of all subleading operators (i.e.
those with naive dimension $\Delta_0 < 6$) together with their
conformal descendants (derivatives). In particular one has to
subtract the Konishi singlet ${\cal K}_1$ ($\Delta_0=2$), as well
as all singlet scalar operators of naive dimension $\Delta_0=4$.
All these operators have non-vanishing anomalous dimensions, thus
the coefficients of the subtractions implicit in the notation of
eq.~(\ref{OPE_T}) will depend on the coupling constant $g$. This
technical complication is, however, largely compensated by the
following nice property inherent in the OPE
definition~(\ref{OPE_T}). We do not have to know the explicit
mixing of the (naively purely scalar) operator ${\cal T}$ with the
operators containing fermions, $F_{\mu \nu}$ and derivatives,
since the triple OPE of eq.~(\ref{OPE_T}) embodies them implicitly
(at least up to order $g^2$ which is relevant for the calculation
of the order $g^4$ correction to the anomalous dimension of
$\widehat{\cal T}(x)$). It should be noted that  a similar compact
definition can be given also for the protected  operator ${\cal
D}_{{\bf 20}^\prime}$ discussed in Section~\ref{composite},
namely~\cite{Eden:2001ec} \be \widehat{\cal D}^{ij}_{{\bf
20}^\prime}(x) = {\rm OPE} \left(
 {\cal Q}^{ik}(x+\epsilon)  \ {\cal Q}^{jk}(x-\epsilon)
\right)   \vert_{ \ {\bf 20^{\prime}}, \ \Delta_0=4} \ \ .
\label{OPE_D} \ee It is instructive to verify that this definition
is equivalent to eq.~(\ref{defD_3}) up to order $g^2$.

Let us stress that eq.~(\ref{OQQQ}) implies not only the vanishing
of the logarithmically divergent piece (which is related to the
anomalous dimension), but also of the associated finite part. This
allows us to prove the following Theorem.

Suppose there exists some $SU(4)$ singlet operator of naive
dimension $\Delta_0=6$, ${\widetilde {\cal O}_S}(x_1)$, with
non-vanishing tree-level 2-point function with ${\cal T}$, \ie\
such that
\begin{equation}
\langle {\widetilde {\cal O}_S}(x_1) \ {\cal T}(x_2)
\rangle \vert_{0} \ \neq \ 0 \, .
\label{OsT}
\end{equation}
Suppose also that at order $g^4$ the divergent part of its 4-point
function with three ${\cal Q}^{ij}$'s vanishes
\begin{equation}
\langle {\widetilde {\cal O}_S}(x_1) \ {\cal Q}^{ij}(x_2) \
{\cal Q}^{ik}(x_3) \ {\cal Q}^{jk}(x_4)
\rangle\vert_{g^4,\ {\rm log}} \ = \ 0 \, ,\label{OsQQQ}
\end{equation}
then it follows that the order $g^4$ anomalous dimension of
$\widehat {\cal T}(x)$ is zero.

In fact, eq.~(\ref{OQQQ}) implies that any singlet scalar
$\Delta_0=6$ operator different from $\widehat{\cal T}$ appearing
in the OPE of three ${\cal Q}^{ij}$ must be multiplied by at least
a factor $g^3$. Hence it cannot give logarithmic corrections at
order $g^4$. Thus the product of three ${\cal Q}^{ij}$'s in
eq.~(\ref{OsQQQ}) acts as a projector on $\widehat {\cal T}(x)$,
which we already know has a vanishing order $g^2$ anomalous
dimension $\gamma_1^{\cal T }  = 0 $, hence the function in
eq.~(\ref{OsQQQ}) must be proportional to
\begin{equation}
\gamma_2^{\cal T }\cdot\langle {\cal T}(x_1) \ {\cal Q}^{ij}(x_2)
\ {\cal Q}^{ik}(x_3) \ {\cal Q}^{jk}(x_4) \rangle \vert_{0}  \, .
\end{equation}
This tree-level 4-point function is indeed non-vanishing by the
very definition of $\widehat {\cal T}$, so we conclude that
\begin{equation}
\gamma_2^{\cal T } \ = \ 0\, .
\label{Gamma2T}
\end{equation}
To complete the proof we have to find an operator ${\widetilde{\cal O}_S}$
with the properties~(\ref{OsT}) and~(\ref{OsQQQ}). The details of this very
long calculation will be presented elsewhere~\cite{longT}. Here let us only
note that a possible choice for it is the triple trace totally colour
symmetric and purely scalar operator
\begin{equation}
{\cal O}_S(x)= {\cal O}_{13} + {3 \over 4 } {\cal O}_{14}+ {1
\over 8 }{\cal O}_{15}\, . \label{defOs}
\end{equation}
The motivation for this choice is that when inserted in the
4-point function~(\ref{OsQQQ}) ${\widetilde {\cal O}_S}$ coincides
with ${\cal O}_S$ so one can avoid all the complications stemming
from $g$ dependent subtractions. As a last remark we note that
with this choice for ${\cal O}_S$ the whole function (not only its
divergent part) in eq.~(\ref{OsQQQ}) is zero, because one can
express it in terms of the order $g^2$ and $g^4$ corrections to
2-point and 3-point functions of only protected operators, like
$\langle {\cal Q}^{ij} \ {\cal Q}^{ij} \rangle $ and $\langle
{\cal D}^{ij} \ {\cal Q}^{ik} \ {\cal Q}^{jk} \rangle $, each of
which vanishes.

\section{Conclusions and summary}
\label{concsumm}

Let us summarize our results. We have reexamined the issue of operator
mixing in ${\cal N} = 4$ SYM and argued that particular care should be
exerted in defining regularized composite (gauge invariant) operators, as soon
as one is willing to go beyond one-loop in perturbation theory.

Exact superconformal invariance puts stringent constraints that
imply the resummation of logarithms in power-like behaviours and
allows us to construct a systematic procedure for the explicit
(numerical) resolution of the operator mixing problem and the
calculation of anomalous dimensions.

Our strategy is in a sense complementary to the one advocated
in~ref.~\cite{Beisert:2003tq,Beisert:2003jj} that emphasizes, instead, the
role of the dilatation operator, possibly viewed as the Hamiltonian of a
super spin chain~\cite{Minahan:2002ve,Beisert:2003yb,Beisert:2003ys}.
Although the latter approach seems very efficient for one-loop computations
in ``closed sectors'', we can't help spending a word of caution on
the effectiveness of the method beyond one-loop. At any rate the two
approaches should give equivalent results and it is reassuring that they
indeed do so in the limited number of cases where they can be compared.
It would be very interesting to consider and possibly clarify the
role of the generalized ${\cal N} = 4$ Konishi
anomaly~\cite{Konishi:1983hf,Amati:ft,Cachazo:2002ry} in the mixing of
operators involving fermion ``impurities'' in the approach of
ref.~\cite{Beisert:2003tq,Beisert:2003jj} in view of the presence of odd
powers of $g$ in the expansion of the dilatation operator, even in the
BMN limit.

More importantly, it would be nice if one could resolve the order
$g^4$ mixing for the singlet operators of naive dimension
$\Delta_0=6$ using the dilatation operator
method~\cite{Beisert:2003tq,Beisert:2003jj} and confirm our
surprising result of the vanishing of the anomalous dimension of
the (purely scalar) operator ${\cal{T}}$ up to $g^4$ and possibly
beyond. Understanding whether this is an accident of the low
orders in the perturbative expansion, that is an exact but
isolated case, or rather the first instance of a class of
kinematically unprotected but yet dynamically ``unrenormalized''
operators is a challenge for any future investigation in this
field, as it might point towards some hidden dynamical symmetry
that could be at the heart of the conjectured integrability of
${\cal N} = 4$
SYM~\cite{Minahan:2002ve,Beisert:2003tq,Beisert:2003yb,Klose:2003qc}.
 The story of non-renormalization theorems in ${\cal N} = 4$
SYM theory, \eg\ for extremal
correlators~\cite{D'Hoker:1999ea,Bianchi:1999ie,Eden:1999kw,Eden:2000gg},
is suggestive in this respect. The holographic correspondence,
after the subtle issues of renormalization of composite
operators~\cite{Bianchi:2001de,Bianchi:2001kw,Muck:cm} has been
carefully taken care of, could provide a guide to the
understanding of our puzzling result. Instanton
calculus~\cite{Bianchi:1998nk,Bianchi:1998xk,Green:2002vf,Kovacs:2003rt}
can give further insights into this issue or even lead to a
non-vanishing contribution to $\gamma_{\cal T}$ at the
non-perturbative level, much as it happens for some non-local
observables~\cite{Bianchi:2001jg,Bianchi:2002gz}. \vskip .3cm

{\bf Acknowledgements }

We wish to thank B.~Eden for an early participation to the present
work. Discussions with  N.~Beisert, N.~Constable, M.~D'Alessandro,
B.~Eden, S.~Ferrara, G.~Ferretti, D.~Freedman, L.~Genovese,
U.~Gursoy, K.~Konishi,   S.~Kovacs, J.~F.~Morales, E.~Sokatchev
and O.~Zapata   are gratefully acknowledged. This work was
supported in part by INFN, the EC contract HPRN-CT-2000-00122, the
EC contract HPRN-CT-2000-00148, the EC contract
HPRN-CT-2000-00131, the MIUR-COFIN contract 2001-025492, the INTAS
contract N~2000-254, and the NATO contract PST.CLG.978785.

\end{document}